\documentclass[preprint,showpacs,preprintnumbers,amsmath,amssymb]{revtex4}

\usepackage{graphicx}
\usepackage{dcolumn}
\usepackage{bm}
\usepackage{epsfig}
\newcommand{\be}{\begin{eqnarray}}
\newcommand{\ee}{\end{eqnarray}}

\begin{document}
\title{Magnetism of one-dimensional strongly repulsive spin-$1$ bosons with antiferromagnetic spin exchange interaction}

\author{J.Y. Lee$^{1}$, X.W. Guan$^{1}$,
 M.T. Batchelor$^{1, 2}$, and C. Lee$^{3}$}

 \affiliation{$^{1}$ Department of Theoretical Physics,
 Research School of Physics and Engineering,
 Australian National University, Canberra ACT 0200, Australia}

 \affiliation{$^{2}$ Mathematical Sciences Institute,
 Australian National University, Canberra ACT 0200, Australia}

 \affiliation{$^{3}$ Nonlinear Physics Centre and
 ARC Centre of Excellence for Quantum-Atom Optics,
 Research School of Physics and Engineering, Australian
 National University, Canberra ACT 0200, Australia}

\date{\today}

\begin{abstract}
We investigate magnetism and quantum phase transitions in a
one-dimensional system of integrable spin-$1$ bosons with strongly
repulsive density-density interaction and antiferromagnetic spin
exchange interaction via the thermodynamic Bethe ansatz method.
At zero temperature, the system exhibits three quantum phases: (i) a
singlet phase of boson pairs when the external magnetic field $H$ is
less than the lower critical field $H_{c1}$; (ii) a ferromagnetic
phase of atoms in the hyperfine state $\left|F=1, m_{F}=1
\right\rangle$ when the external magnetic field exceeds the upper
critical field $H_{c2}$; and (iii) a mixed phase of singlet pairs
and unpaired atoms in the intermediate region $H_{c1}<H<H_{c2}$.
At finite temperatures, the spin fluctuations affect the
thermodynamics of the model through coupling the spin bound states
to the dressed energy for the unpaired $m_{F}=1$ bosons.
However, such spin dynamics is suppressed by a
sufficiently strong external field at low temperatures.
Thus the singlet pairs and unpaired bosons may form a two-component
Luttinger liquid in the strong coupling regime.

\end{abstract}



\maketitle

\section{Introduction}
The well-developed techniques for controlling and manipulating
Bose-Einstein condensates (BECs) of  spinor atoms provide an
excellent opportunity to explore novel magnetism and quantum phases.
In a magnetic trap, atoms with  different magnetic moments are
subjected to different forces, so it is very difficult to confine a
spinor BEC involving all possible spin states. However, since the
laser-atom interaction is determined by the induced electric dipole
moment, an optical trap may confine every spin state to preserve the
``vector'' property of spinor atoms. This allows one to trap a true
spinor BEC, which involves an ensemble of Bose atoms condensed in a
coherent superposition of all possible hyperfine states. In this
way, several experimental groups have successfully demonstrated
spinor BECs of $^{23}$Na \cite{Stamper-Kurn1998,Miesner1999} and
$^{87}$Rb \cite{Matthews1999, Barrett2001} atoms.

The ground states and some low-energy excitations of a spinor BEC
were theoretically analyzed by Ohmi and Machida \cite{Ohmi1998} and
Ho \cite{Ho1998}.
It has been shown experimentally that all three spin components of a
spin-1 condensate can be either miscible or immiscible with one
another where the immiscibility will lead to formation of spin
domains \cite{Stenger1998}. This has been confirmed numerically
using the Gross-Pitaevskii equation and Thomas-Fermi approximation
\cite{Isoshima1999}. Using the single-mode approximation (SMA), the
ground state population dynamics of a spin-1 BEC have been studied
by Law \emph{et al.} \cite{Law1998} and Pu \emph{et al.}
\cite{Pu1999}.
They have also found that the ground state is a superposition of
collective spin states (Fock states) and cannot be expressed as a
product of individual spin states. This shows the collective
behavior of all three spin components. Ho and Yip \cite{Ho2000},
found the antiferromagnetic ground state to be a fragmented
condensate with large particle number fluctuations as stated in the
references \cite{Law1998,Pu1999}. This fragmented condensate
gradually deforms into a more stable coherent state as the strength
of the external field gradient increases. Recently, Rizzi \emph{et
al.} \cite{Rizzi2008} applied the DMRG method to determine the phase
diagram for one-dimensional spin-1 bosons. In accordance with
Imambekov \emph{et al.} \cite{Imambekov2003} and Yip \cite{Yip2003},
they showed that the dimerized state is among the ground states. The
quantum phases such as the polar phase, nematic phase and spin
singlet phase were discussed by Demler and Zhou \cite{Demler2002}.

In one dimension (1D), spinor Bose gases have a ferromagnetic ground
state in the absence of spin-dependent forces
\cite{Lieb,Yang-Li,Guan2007}. However, the spinor Bose gas can have
either a ferromagnetic or an antiferromagnetic ground state in the
presence of spin-exchange interaction \cite{Ho1998,Ho2000,Ueda}.
Very recently, Cao \emph{et al.} \cite{Cao2007} proved that there
exists an integrable point in scattering parameter space for 1D
spin-1 bosons with both delta-function contact interaction and
spin-exchange interaction. This model provides an important
benchmark to understand spinor BECs and spin liquids in low
dimensions. From the exact Bethe ansatz (BA) solution, Cao \emph{et
al.} found that the ground state is a spin singlet in the absence of
an external field. Essler \emph{et al.} \cite{Essler2009} then
proved that the low energy physics in the weak repulsive coupling
regime can be described by a spin-charge separated theory of an
effective Tomonaga-Luttinger Hamiltonian and an $O(3)$ nonlinear
sigma model. In this weak coupling limit, both the collective
pairing fluctuations and spin fluctuations dominate the low-lying
excitations. Using the BA equations and the effective field theory,
Essler \emph{et al.} calculated the scaling dimensions and the
large-distance asymptotics of correlation functions of the model.
Such a spin liquid phase was also previously investigated by Zhou
\cite{Zhou2001} through the introduction of the Weyl representation
of $SU(2)$.

In this paper, we investigate quantum liquid phases in the 1D
integrable system of spin-1 bosons \cite{Cao2007} with strongly
repulsive and antiferromagnetic spin-exchange interactions. We
derive the thermodynamic Bethe ansatz (TBA) equations on the basis
of particle-hole excitations \cite{Yang1969} and spin-strings
\cite{Takahashi1971}. From these equations we find that for the
strong coupling regime there is a large energy gap in the lowest
spin excitation. We also show that spin fluctuations are frozen out
under a strong external field at zero temperature. When the external
magnetic field exceeds the lower critical field $H_{c1}$, the energy
gap vanishes and the charge excitations evolve into two gapless
modes of singlet pairs and the branch of magnetic quantum number
$m_F=1$ atoms. The external field may break a singlet pair into two
unpaired atoms of $m_{F}=1$ under a sufficiently strong magnetic
field. A ferromagnetic Tonks-Girardeau gas of $m_F=1$ atoms appears
if the external field exceeds the upper critical field $H_{c2}$. The
singlet pairs and unpaired $m_{F}=1$ bosons coexist in the
intermediate region $H_{c1}<H<H_{c2}$.  We show that for strong
coupling the low energy physics of the gapless phase is described by
the universality class a two-component Luttinger liquid as long as
the spin dynamics are frozen out. Moreover, from the TBA equations,
we obtain exact results for the ground state energy and magnetism
for the system with an external magnetic field, which provide an
exact phase diagram and the universality class of quantum phase
transitions for the integrable spinor $F=1$ Bose gas with strongly
repulsive and antiferromagnetic spin-exchange interactions.

\section{The model}
We consider $N$ particles confined in 1D to a length $L$ with
delta-function type density-density interactions and spin-spin
interactions between two atoms. In first quantized form, the
Hamiltonian of this model is given by \cite{Ho1998}
\begin{equation}
H = -\frac{\hbar^{2}}{2m}\sum_{i=1}^{N}\frac{\partial^{2}}{\partial
x_{i}^{2}}+\sum_{i<j}[c_{0}+c_{2}\mathbf{F}_{i}\cdot\mathbf{F}_{j}]\delta(x_{i}-x_{j})
+E_{z}\label{Ham}
\end{equation}
where $\mathbf{F}_{i}$ are spin-1 operators,
$c_{0}=(g_{0}+2g_{2})/3$ and $c_{2}=(g_{2}-g_{0})/3$ are the
interaction parameters which are related to the $s$-wave scattering
lengths $a_{F}$ in the spin-0 and spin-2 channels, and
$g_{F}=4\pi\hbar^{2}a_{F}/m$ with $m$ being the mass of the atoms.
The term $E_{z}$ accounts for the Zeeman energy which will later be
given explicitly. Throughout this paper we use the dimensionless
units of $\hbar=2m=1$ for convenience. We are interested in the
antiferromagnetic case $c_{2}>0$ and $c_{0}=c_{2}=c$ where this
model is exactly solvable by the BA \cite{Cao2007}.  The model that
we examine also has repulsive density-density interactions since the
interaction parameter $c_{0}=c_{2}>0$.  Repulsive interactions
result in an effective attraction in the two-body scattering matrix
for the spin-0 channel and an effective repulsion in the scattering
matrix for the spin-2 channel. This naturally leads to the formation
of singlet bound pairs in the spin-0 channel \cite{Cao2007}. Due to
the existence of the spin exchange interaction in the Hamiltonian,
the number of particles in a particular spin state ($m_{F}=1,0,-1$)
is no longer conserved because spin transmutation is allowed to
occur. The scattering between two particles of spin $m_{F}=1$ and
$m_{F}=-1$ can produce two particles of spin $m_{F}=0$ and
vice-versa. The only conserved quantities are the total particle
number $N$ and the total spin in the $z$-component $S^{z}$. This
model possesses $U(1)$ symmetry for charge conservation and $SU(2)$
symmetry which corresponds to spin conservation. For weak
interaction and in the absence of an external field, the spin
dynamics is described by the $O(3)$ non-linear sigma model which can
be separated from the BA equations \cite{Essler2009}. The charge
sector on the other hand is described by collective pairing density
fluctuations of free boson fields.

The BA equations for this Hamiltonian acting on a totally symmetric
Bose wavefunction are \cite{Cao2007}
\begin{eqnarray}
\exp(\mathrm{i}k_{j}L) &=& -\prod_{l=1}^{N}e_{4}(k_{j}-k_{l})
\prod_{\alpha=1}^{M}e_{-2}(k_{j}-\Lambda_{\alpha}),\nonumber\\
\prod_{l=1}^{N}e_{2}(\Lambda_{\alpha}-k_{l}) &=&
-\prod_{\beta=1}^{M}e_{2}(\Lambda_{\alpha}-\Lambda_{\beta}),
\label{BA}
\end{eqnarray}
where $j=1,\ldots,N$ and $\alpha=1,\ldots,M$. $\{k_{j}\}$ is the set
of quasimomenta for the particles and $\{\Lambda_{\alpha}\}$ is the
set of spin rapidities that characterize the internal spin degrees
of freedom. The quantum number $M$ is a conserved quantity
satisfying the relation $M=N-S^{z}$ and the function
\begin{equation}
e_{n}(x)=\frac{x+\mathrm{i}nc'}{x-\mathrm{i}nc'}
\end{equation}
where $c'=c/4$. The energy $E=\sum_{j}k_{j}^{2}$ and total momentum
$p=\sum_{j}k_{j}$ of the system can be obtained by solving the
coupled BA equations for the sets $\{k_{j}\}$ and
$\{\Lambda_{\alpha}\}$.

\section{The TBA equations}
In the thermodynamic limit $N,L\rightarrow\infty$ with the ratio
$N/L$ finite, the sets of solutions $\{k_{j}\}$ and
$\{\Lambda_{\alpha}\}$ of the BA equations take certain forms. As
mentioned in ref. \cite{Cao2007}, the $k_{j}$'s and
$\Lambda_{\alpha}$'s can form complex pairs
$k_{j}=\lambda_{j}\pm\mathrm{i}c'$ and
$\Lambda_{j}=\lambda_{j}\pm\mathrm{i}c'$ where $\lambda_{j}$ is
real. In Fig. 1, we show a schematic configuration of the
quasimomenta and spin rapidities for the ground state. Notice that
each pair of $k_{j}$'s share the same real part as a corresponding
pair of $\Lambda_{\alpha}$'s. The bound states are associated with a
pair of $m_F=\pm1$ bosons or two $m_F=0$ bosons. In the absence of
an external field, this bound state is created by the operator
$A^{\dagger}\equiv
[(a_0^{\dagger})^2-2a^{\dagger}_1a^{\dagger}_{-1}]/\sqrt{3}$
\cite{Ueda}. In the presence of a sufficiently strong magnetic
field, the singlet bound state of two $m_{F}=0$ bosons is not
energetically favored \cite{Ho2000}. In addition to that, we also
have real $k_{j}$'s and $\Lambda$-strings of the form
$\Lambda_{\alpha}^{n,j}=\Lambda_{\alpha}^{n}+\mathrm{i}(n+1-2j)c'$,
$j=1,\ldots,n$ as solutions. The spin-strings characterize the spin
wave fluctuations. In the thermodynamic limit, the grand partition
function is $Z=tr(\mathrm{e}^{-H/T})=\mathrm{e}^{-G/T}$
\cite{Yang1969,Takahashi1999} where the Gibbs free energy $G=E
+E_{\rm Z}-\mu N-TS$, chemical potential $\mu$, Zeeman energy
$E_{\rm Z}=-HS^z$ and entropy $S$ are given in terms of the
densities of charge bound states and spin-strings which are subject
to the BA equations (\ref{BA}).

The equilibrium states are determined by minimizing the Gibbs free
energy, which gives rise to a set of coupled nonlinear integral
equations -- the TBA equations, i.e.
\begin{eqnarray}
\varepsilon_{1}(k) &=&
k^{2}-\mu-H-Ta_{4}\ast\ln\left(1+e^{-\varepsilon_{1}(k)/T}\right)\nonumber\\
&&+T[a_{1}-a_{5}]\ast\ln\left(1+e^{-\varepsilon_{2}(k)/T}\right)
\nonumber\\
&&-T\sum_{n=1}^{\infty}[a_{n-1}+a_{n+1}]\ast\ln\left(1+e^{-\phi_{n}(k)/T}\right),\nonumber\\
\varepsilon_{2}(k) &=&
2(k^{2}-c'^{2}-\mu)+T[a_{1}-a_{5}]\ast\ln\left(1+e^{-\varepsilon_{1}(k)/T}\right)\nonumber
\\
&&
+T[a_{2}-a_{4}-a_{6}]\ast\ln\left(1+e^{-\varepsilon_{2}(k)/T}\right),\nonumber
\\
 \phi_{n}(k)&=&nH+T[a_{n-1}+a_{n+1}]\ast\ln\left(1+e^{-\varepsilon_{1}(k)/T}\right)\nonumber\\
&&+T\sum_{m=1}^{\infty}T_{mn}\ast\ln\left(1+e^{-\phi_{n}(k)/T}\right).\label{TBA}
\end{eqnarray}
We present a detailed derivation of the TBA equations in the
Appendix. In the above equations, the convolution $f\ast g(x)$ and
the functions $a_{n}(x)$ are defined in equations
(\ref{Convolution}) and (\ref{a}). The function $T_{mn}(x)$ is also
given in the Appendix.  The TBA equations are expressed in terms of
the dressed energies $\varepsilon_{1}(k)$, $\varepsilon_{2}(k)$ and
$\phi_{n}(k)$ for unpaired states, paired states and spin-strings,
respectively. The dressed energy $\varepsilon_{1}(k)$ depends not
only on the chemical potential $\mu$ and the external field $H$ but
also on the interactions between unpaired bosons and singlet pairs
as well as the spin fluctuations characterized by the spin-strings
$\phi_{n}(k)$.  Physically, the dressed energies measure the
energies over the ``Fermi surfaces''. We clearly see that spin
fluctuations are coupled to the dressed energy of unpaired $m_{F}=1$
bosons $\varepsilon_{1}(k)$ through the last term in the first
equation in (\ref{TBA}). The spin flippings caused by thermal
fluctuations are described by the last equation where the magnon
excitations in $m_{F}=1$ bosons are described by an effective
ferromagnetic spin-spin interaction. There is no such spin
fluctuation coupled to the dressed energy of bound pairs due to its
spin neutral effect.

In the strong coupling limit, the dressed energies
$\varepsilon_{1}(k)$ and $\varepsilon_{2}(k)$ marginally depend on
the pressures of each other, see the dressed energy-dependent terms
in the first and second equations. This is similar to the
configuration for the attractive Fermi gas
\cite{GBLB,Iida2007,ZGLBO}. The only difference is that here the
unpaired bosons may scatter between themselves whereas in the
attractive Fermi gas unpaired fermions do not scatter among
themselves. The pressure per unit length of the system is derived
from the expression $p=-\partial G/\partial L$ as
\begin{eqnarray}
p&=&\frac{T}{2\pi}\int_{-\infty}^{\infty}\ln\left(1+e^{-\varepsilon_{1}(k)/T}\right)dk\nonumber\\
&&+\frac{T}{\pi}\int_{-\infty}^{\infty}\ln\left(1+e^{-\varepsilon_{2}(k)/T}\right)dk,
\end{eqnarray}
where the first term corresponds to the pressure for  unpaired
bosons and the second term to the pressure for singlet pairs.

\section{Quantum phase transitions and magnetism}
Solving the TBA equations (\ref{TBA}) imposes a formidable challenge
due to the involvement of infinitely many spin-strings. Here we
shall focus on the ground state properties and quantum phase
transitions driven by an external magnetic field. Following the
method developed in \cite{GBLB,HFGB}, we state two conditions to
proceed on, namely we consider: (I) the scenario where we are in the
ground state with $T\rightarrow 0$, and (II) the strong coupling
limit $c\gg 1$. With these two conditions, we can obtain a series
expansion in terms of the coupling strength $1/c$ for various
thermodynamic quantities, as we shall see later. The strong
interaction condition should be easily reached because generally the
interaction energy is much larger than the kinetic energy for a
dilute gas in 1D with finite interaction strength $c$.

When $T\rightarrow 0$, the TBA equations (\ref{TBA})  simplify to
\begin{eqnarray}
\varepsilon_{1}(k)&=&k^{2}-\mu-H+a_{4}\ast\varepsilon_{1}^{-}(k)
+[a_{5}-a_{1}]\ast\varepsilon_{2}^{-}(k)\nonumber \\
\varepsilon_{2}(k)&=&2(k^{2}-c'^{2}-\mu)+[a_{5}-a_{1}]\ast\varepsilon_{1}^{-}(k)\nonumber \\
&&+[a_{6}+a_{4}-a_{2}]
\ast\varepsilon_{2}^{-}(k),\label{TBA-d}
\end{eqnarray}
where the dressed energies $\varepsilon^{-}_a(k)$ with $a=1,\,2$
imply that we only consider the domain where the function
$\varepsilon_{a}(k)<0$. The negative part of the dressed energies
$\varepsilon_{a}(k)$ for $k\le Q_{a}$ corresponds to occupied states
in the dressed energies while the positive part of $\varepsilon_{a}$
corresponds to unoccupied states. The integration boundaries $Q_{a}$
characterize the ``Fermi surfaces'' at $\varepsilon_a(Q_{a})=0$.
There are no $\Lambda$-strings involved in the ground state (all
$\phi_n(k)$ are not occupied), thus the dressed energy equations
evolve into two coupled dressed energies. This characterizes the
scattering among singlet bound pairs and unpaired bosons. They
provide complete phase diagrams and information of quantum phase
transitions with respect to the Zeeman splitting parameter $H$ and
the chemical potential $\mu$. The pressure of the system can be
represented in a neater way if we introduce the following notation
when $T\rightarrow 0$:
\begin{eqnarray}
\nonumber p &=&
-\frac{1}{2\pi}\int_{-\infty}^{\infty}\varepsilon_{1}^{-}(k)dk-\frac{1}{\pi}\int_{-\infty}^{\infty}\varepsilon_{2}^{-}(k)dk
\\ &\equiv& p_{1}+p_{2}.
\end{eqnarray}
Every thermodynamic quantity with a subscript 1 (or 2) corresponds
to unpaired states (or paired states).

When $c\gg 1$, we can take a Taylor expansion of the functions
$a_{n}$. Throughout this paper, we only keep track of terms up to
order $1/c^{2}$. Higher order corrections can be calculated in a
straightforward manner. In this limit, equation (\ref{TBA-d})
becomes (up to order $1/c^2$)
\begin{eqnarray}
\nonumber \varepsilon_{1}(k)&\approx&
k^{2}-\mu-H-\frac{p_{1}}{2c'}+\frac{4p_{2}}{5c'},\\
\varepsilon_{2}(k)&\approx &2(k^{2}-c'^{2}-\mu)+\frac{8p_{1}}{5c'}+\frac{p_{2}}{12c'}.\label{T-E}
\end{eqnarray}
We then integrate equations (\ref{T-E}) between the ``Fermi points''
$\pm Q_{1}$ and $\pm Q_{2}$ so that we can re-write the equations in
terms of $p_{1}$ and $p_{2}$. This gives
\begin{eqnarray}
-2\pi p_{1} &\approx & \frac{2}{3}Q_{1}^{3}-2\mu Q_{1}-2H
Q_{1}-\frac{p_{1}Q_{1}}{c'}+\frac{8p_{2}Q_{1}}{5c'},\nonumber\\
-\pi p_{2}&\approx &\frac{4}{3}Q_{2}^{3}-4c'^{2}Q_{2}-4\mu
Q_{2}+\frac{16p_{1}Q_{2}}{5c'}+\frac{p_{2}Q_{2}}{6c'}.\label{T-P}
\end{eqnarray}
We also make use of the fact that the dressed energies
$\varepsilon_{1}(k)$ and $\varepsilon_{2}(k)$ vanish at the ``Fermi
points'' i.e., $\varepsilon_{1}(\pm Q_{1})=0$ and
$\varepsilon_{2}(\pm Q_{2})=0$,
\begin{eqnarray}
Q_{1}^{2}&\approx &\mu+H+\frac{p_{1}}{2c'}-\frac{4p_{2}}{5c'},\nonumber\\
Q_{2}^{2}&\approx &\mu+c'^{2}-\frac{4p_{1}}{5c'}-\frac{p_{2}}{24c'}.
\end{eqnarray}

Substituting the ``Fermi points'' into equations (\ref{T-P}) and
then re-arranging and iterating the terms yield
\begin{eqnarray}
p_{1}&\approx &\frac{2\mu_{1}^{3/2}}{3\pi}+\frac{\mu_{1}^{3/2}}{4\pi}\left(\frac{p_{1}}{2\mu_{1}c'}
-\frac{4p_{2}}{5\mu_{1}c'}\right)^{2}\nonumber\\
&&
+\frac{\mu_{1}^{3/2}}{\pi}\left(\frac{p_{1}}{2\mu_{1}c'}
-\frac{4p_{2}}{5\mu_{1}c'}\right),\nonumber\\
p_{2}&\approx &\frac{8\mu_{2}^{3/2}}{3\pi}+\frac{\mu_{2}^{3/2}}{\pi}\left(\frac{4p_{1}}{5\mu_{2}c'}
+\frac{p_{2}}{24\mu_{2}c'}\right)^{2}\nonumber\\
&& -\frac{4\mu_{2}^{3/2}}{\pi}\left(\frac{4p_{1}}{5\mu_{2}c'}
+\frac{p_{2}}{24\mu_{2}c'}\right),
\end{eqnarray}
where we denote the effective chemical potentials
$\mu_{1}\equiv\mu+H$ and $\mu_{2}\equiv\mu+c'^{2}$ for the unpaired
and paired bosons. From the relations $n=\partial p/\partial\mu$ and
$nm^{z}=\partial p/\partial H$ and after some lengthy iterations, we
arrive at the expressions for the chemical potentials of unpaired
and paired bosons,
\begin{eqnarray}
\mu_{1}&\approx &
\pi^{2}n^{2}\left[(m_{z})^{2}\left(1-\frac{16m^{z}}{3\gamma}+\frac{32(1-m^{z})}{5\gamma}\right)\right.\nonumber\\
&&\left.+\frac{2(1-m^{z})^{3}}{15\gamma}\right],\nonumber
\\
\mu_{2}&\approx &\frac{\pi^{2}n^{2}}{16}\left[(1-m^{z})^{2}\left(1+
\frac{4(1-m^{z})}{9\gamma}+\frac{32m^{z}}{5\gamma}\right)\right.\nonumber
\\
&&\left. +\frac{512(m^{z})^{3}}{15\gamma} \right],\label{mu}
\end{eqnarray}
where $\gamma=c/n$. Substituting these two equations back into
$p_{1}$ and $p_{2}$ gives the pressures
\begin{eqnarray}
p_{1}&\approx &\frac{2}{3}\pi^{2}n^{3}(m^{z})^{3}\left(1-\frac{6m^{z}}{\gamma}+\frac{48(1-m^{z})}{5\gamma}\right),\nonumber\\
p_{2}&\approx &\frac{1}{24}\pi^{2}n^{3}(1-m^{z})^{3}\left(1+\frac{(1-m^{z})}{2\gamma}+\frac{48m^{z}}{5\gamma}\right).
\end{eqnarray}
Further, the free energy can be obtained as
\begin{eqnarray}
\nonumber F  &\approx & \nonumber\frac{1}{3}\pi^{2}n^{3}(m^{z})^{3}\left(1-\frac{4m^{z}}{3\gamma}+\frac{32(1-m^{z})}{5\gamma}\right)\nonumber\\
&&+\frac{1}{48}\pi^{2}n^{3}(1-m^{z})^{3}\left(1+\frac{(1-m^{z})}{3\gamma}+\frac{32m^{z}}{5\gamma}\right)\nonumber\\
&&
-\frac{c^{2}}{16}n(1-m^{z})-Hnm^{z}+O\left(\frac{1}{\gamma^{2}}\right).
\end{eqnarray}

To find the ground state energy, we can use the relation
$E=F+Hnm^{z}$. There is also an alternative way to derive the ground
state energy based on the definition $E=\sum_{j}k_{j}^{2}$ and the
distribution of $\{k_{j}\}$ in quasimomenta space. Indeed, we show
that the energy per unit length derived from the discrete BA
equations (\ref{BA}) for arbitrary magnetization,
\begin{eqnarray}
\nonumber \frac{E}{L} &=&
\frac{1}{3}\pi^{2}n_{1}^{3}\left(1+\frac{2(32n_{2}-10n_{1})}{5c}+\frac{3(32n_{2}-10n_{1})^{2}}{25c^{2}}\right)\nonumber
\\ && +\frac{1}{6}\pi^{2}n_{2}^{3}\left(1+\frac{2(48n_{1}+5n_{2})}{15c}+\frac{3(48n_{1}+5n_{2})^{2}}{225c^{2}}\right)\nonumber\\
&&-\frac{n_{2}c^{2}}{8}+O\left(\frac{1}{c^{3}}\right)
\end{eqnarray}
coincides with the TBA result $E=F+Hnm^{z}$ up to the order of $1/c$
through the relations $n_{1}=nm^{z}$ and
$n_{2}=\frac{n}{2}(1-m^{z})$ where $n_{1}$ and $n_{2}$ are the
density of unpaired and paired bosons, respectively. However, the
dressed energy formalism provides a more elegant way to study
quantum phase transitions \cite{BGOT}.

For strong coupling, a pair of two bosons becomes stable because the
binding energy $\epsilon_b=\frac{\hbar^2}{2m}\frac{c^2}{8}$ can
exceed the kinetic energy. Therefore, the ground state in the
absence of an external field is characterized by an empty ``Fermi
sea'' for unpaired bosons and a fully filled ``Fermi sea'' for bound
pairs. From the dressed energy equations (\ref{TBA-d}), we find that
quantum phase transitions driven by an external field can be
determined by the energy transfer relation
$H=\mu_{1}-\mu_{2}+c^{2}/16$, i.e.,
\begin{eqnarray}
&&H \approx
n^{2}\left[\frac{\gamma^{2}}{16}+\pi^{2}(m^{z})^{2}\left(1-\frac{112m^{z}}{15\gamma}
+\frac{32(1-m^{z})}{5\gamma}\right.\right.
\nonumber \\
&& \left.+\frac{164(m^{z})^{2}}{5\gamma^{2}}-\frac{1792m^{z}(1-m^{z})}{25\gamma^{2}}
+\frac{768(1-m^{z})^{2}}{25\gamma^{2}}\right) \nonumber \\
&&
-\frac{\pi^{2}(1-m^{z})^{2}}{16}\left(1-\frac{76(1-m^{z})}{45\gamma}+\frac{32m^{z}}{5\gamma}+\frac{768(m^{z})^{2}}{25\gamma^{2}}\right. \nonumber \\ &&
\left.\left.-\frac{167(1-m^{z})^{2}}
{180\gamma^{2}}-\frac{1216m^{z}(1-m^{z})}{75\gamma^{2}}\right)\right]\label{H-E}
\end{eqnarray}
where we have used the relations
$\mu_{\kappa}=\frac{\partial}{\partial
n_{\kappa}}\left(E/L+n_{2}\epsilon_{b}\right)$ for $\kappa=1,2$ to
obtain expressions for the chemical potentials up to order
$1/\gamma^{2}$.

The lower critical field $H_{c1}$ diminishes the gap, thus a phase
transition from a singlet ground state into a gapless phase, where
two dressed energies of the paired and unpaired bosons couple to
each other, occurs when $H>H_{c1}$. When the external field exceeds
the upper critical field $H_{c2}$, all singlet bound pairs are
broken which leads to a ferromagnetic Tonks-Girardeau Bose gas.  The
lower and upper critical fields are found by letting $m^{z}=0$ and
$m^{z}=1$ in (\ref{H-E}), with result
\begin{eqnarray}
H_{c1}&\approx &\frac{n^{2}}{16}\left[\gamma^{2}-\pi^{2}\left(1-\frac{76}{45\gamma}-\frac{167}{180\gamma^{2}}\right)
\right],\nonumber\\
H_{c2}&\approx &\frac{n^{2}}{16}\left[\gamma^{2}+16\pi^{2}\left(1-\frac{112}{15\gamma}+\frac{164}{5\gamma^{2}}
\right)\right].
\end{eqnarray}
In Fig. 2, we show the magnetization vs external field for different
values of the interaction strength $c$. We see clearly that for
$H<H_{c1}$ there is no breaking of bound pairs.  The magnetization
gradually increases from zero to $n$ as $H$ gradually approaches
$H_{c2}$.  The phase transitions across $H_{c1}$ and $H_{c2}$ are of
second order. In the vicinities of $H_{c1}$ and $H_{c2}$, the
leading order of the respective normalized magnetizations are given
by
\begin{eqnarray}
m_{1}^{z}&\approx &\frac{8(H-H_{c1})}{\pi^{2}n^{2}}\left(1+\frac{86}{15\gamma}-\frac{2813}{450\gamma^{2}}\right),\nonumber\\
m_{2}^{z}&\approx &1-\frac{(H_{c2}-H)}{2\pi^{2}n^{2}}\left(1+\frac{72}{5\gamma}-\frac{2536}{25\gamma^{2}}\right).
\end{eqnarray}
which show a linear dependence on the external field near the
critical points. For an external field $H_{c1}<H<H_{c2}$, the
singlet paired state and unpaired state coexist. They form a
two-component Luttinger liquid in this gapless phase.

In Fig. 3, we show the ground-state phase diagram in the $n-H$
plane. As $n \to 0$, both critical fields approach the same value
$H_c=\epsilon_{\rm b}/2$. The solid (dashed) lines correspond to the
two critical fields for the case $c=20$ ($c=40$). The ferromagnetic
phase of all atoms in state $\left|F=1,m_{F}=1\right\rangle$ appears
above the critical field $H_{c2}$, the singlet phase of singlet
pairs appears below the critical field $H_{c1}$ and the mixed phase
of atoms in state $\left|F=1,m_{F}=1\right\rangle$ and singlet pairs
appears between the two critical fields.

\section{The spin and charge velocities}
In 1D systems, spin-charge separation is the hallmark of many-body
physics \cite{Giamarchi2003}. The collective charge excitations are
described by sound modes with a linear dispersion.
The spin excitations are gapped with a dispersion
$\epsilon_{\nu}(p)=\sqrt{\Delta_{\nu}^2+v_{\nu}^2p^2}$ where
$\Delta_{\nu}$ is the excitation gap and $v_{\nu}$ is the spin
velocity in spin branch $\nu$.
This leads to the phenomenon of spin-charge separation. A method has
been proposed to probe this phenomenon experimentally in a 1D system
of interacting electrons at low energies \cite{Jompol2009}.

To calculate the charge velocity, we need to find the energy of the
lowest excited state that does not involve breaking any pairs. In
the absolute ground state where $H=0$, the system is only made up of
fully paired states below the ``Fermi level'' and the total momentum
of the system is zero. This is achieved when there is no magnetic
field present. To excite the system, we allow the pair with the
largest momentum to leave the ``Fermi sea'' and let the excited
state have a total momentum of $p$, i.e., $\sum_{j}k_{j}=p$. We then
calculate its total energy $E=\sum_{j}k_{j}^{2}$. The difference
between the excitation energy and the ground state energy is equal
to the charge velocity times $p$. The energy difference in the
thermodynamic limit is thus calculated to be
\begin{equation}
E(p)-E_{0}=\pi
n_{2}p\left(1+\frac{2n_{2}}{3c}\right)+O\left(\frac{1}{c^{2}}\right).
\end{equation}
Therefore in terms of the total particle number $n=N/L$ and
interaction strength $c$, the charge velocity is
\begin{equation}
v_{c}=\frac{\pi
n}{2}\left(1+\frac{1}{3\gamma}\right)+O\left(\frac{1}{c^{2}}\right).
\end{equation}

An alternative way to calculate the charge velocity for the singlet
ground state is through the relation
\begin{equation}
v_{c}=\sqrt{\frac{2L}{n}\left(\frac{\partial^{2}E_{0}}{\partial
L^{2}}\right)}.
\end{equation}
Both methods yield the same result.

The spin velocity on the other hand is calculated by considering the
lowest excited state where one pair is broken into two unpaired
states. Both unpaired states will occupy opposite ends of the
momentum distribution so that the excitation energy is minimized.
The total momentum of the excited state can be parameterized by $p$
in the same manner as before. We can equate the energy difference
between the excited state and the fully paired ground state to the
energy dispersion $\epsilon(p)$. In the thermodynamic limit,
\begin{equation}
E(p)-E_{0}=\frac{c^{2}}{8}+\frac{p^{2}}{2}\left(1+\frac{64n_{2}}{5c}\right)+O\left(\frac{1}{c^{2}}\right)
\equiv\epsilon(p).
\end{equation}
From the original relation
$\epsilon(p)=\sqrt{\Delta^{2}+v_{s}^{2}p^{2}}$, we obtain the
relation $\epsilon(p)=\Delta+\frac{v_{s}^{2}p^{2}}{2\Delta}$ in the
limit $\Delta\gg 1$ where the gap is very large. Comparing both
expressions for the dispersion energy, we can easily verify that
$\Delta=c^{2}/8$ and
\begin{equation}
v_{s}^{2}=\frac{c^{2}}{8}\left(1+\frac{64n_{2}}{5c}\right).
\end{equation}
Hence the spin velocity is
\begin{equation}
v_{s}=\frac{c}{2\sqrt{2}}\left(1+\frac{16n}{5c}\right)+O\left(\frac{1}{c^{2}}\right).
\end{equation}
The spin velocity is divergent due to a large energy gap as $c\to
\infty$. This demonstrates that there is spin-charge separation over
the singlet ground state. We also note that this phenomenon depends
on the state of the system within an external field. Essler \emph{et
al.} \cite{Essler2009} showed that spin and charge velocities are
equal in the weak coupling limit when there is no external field
involved.

In the gapless phase when $H_{c1}<H<H_{c2}$, the ground state
($T\rightarrow 0$) of this system is conformally invariant
\cite{Blote1986,Affleck1986}. The excitations close to the ``Fermi
surfaces'' in unpaired and pair branches have linear dispersions.
The finite-size corrections to the ground state energy are given by
\begin{equation}
E_{0}=Le_{0}^{\infty}-\frac{\pi c}{6L}(v_{c}^{(1)}+v_{c}^{(2)})
\end{equation}
where the central charge $c=1$ for this system, $E_{0}$ is the
ground state energy for the finite system and $e_{0}^{\infty}$ is
the ground state energy density for the infinite system. The charge
velocities for unpaired and paired bosons are given explicitly by
the expressions
\begin{eqnarray}
v_{c}^{(1)}&\approx &2\pi
n_{1}\left(1+\frac{2(32n_{2}-10n_{1})}{5c}+\frac{3(32n_{2}-10n_{1})^{2}}{25c^{2}}\right),\nonumber\\
v_{c}^{(2)}&\approx &\pi
n_{2}\left(1+\frac{2(48n_{1}+5n_{2})}{15c}+\frac{3(48n_{1}+5n_{2})^{2}}{225c^{2}}\right).
\end{eqnarray}
In this phase, spin fluctuations are frozen out and thus the  charge
density fluctuations dominate the ground state and is effectively
described by the universality class of a two component Luttinger
liquid.

\section{Conclusion}

In conclusion we derived the TBA equations for a system of 1D spin-1
bosons with repulsive density-density and antiferromagnetic spin
exchange interactions and solved the TBA equations for the zero
temperature case in the strong coupling limit. We obtained the
ground state energy, chemical potentials, critical fields and
magnetization in terms of interaction strength and the external
magnetic field. We also presented an exact phase diagram of strongly
interacting spin-1 bosons which  facilitates experimental analysis
of phase segments. For the weak coupling limit, the collective
excitations in the charge sector is described by a
Tomonaga-Luttinger liquid, whereas the spin dynamics is described by
the $O(3)$ non-linear sigma model \cite{Essler2009}. However, for
the strong coupling limit, spin fluctuations can be suppressed by a
strong external magnetic field. The density fluctuations thus evolve
into a two-component Luttinger liquid. At zero temperature, the
model exhibits three quantum phases: singlet pairs of two bosons for
external field $ H<H_{c1}$; a fully-polarized Tonks-Girardeau gas
phase of $m_F=1$ bosons for $H> H_{c2}$; and a mixed phase of
singlet pairs and unpaired $m_{F}=1$ atoms for an intermediate field
$H_{c1} < H < H_{c2}$. The phase transitions in the vicinities of
$H_{c1}$ and $H_{c2}$ are of second order with a
linear-field-dependent magnetization. Our results provide a new
aspect of this model, namely spin liquid v.s. Luttinger liquid
behavior.

\vspace{1cm}
This work has been supported by the Australian Research Council.
We thank Profs J.-P. Cao, S. Chen and Y.-P. Wang  for helpful
discussions. C.L. thanks Yu.S. Kivshar for support.

\appendix
\section{Derivation of the TBA Equations}

Here we derive the TBA equations in detail following the steps for
the attractive spin-1/2 fermion model in Chapter 13 of Takahashi's
book \cite{Takahashi1999}. Substituting all possible solutions for
$\{k_{j}\}$ and $\{\Lambda_{\alpha}\}$ back into the BA equations
(\ref{BA}) gives
\begin{eqnarray}
\nonumber\exp(\mathrm{i}k_{j}L) &=& \prod_{l\neq
j}^{N_{1}}e_{4}(k_{j}-k_{l})\prod_{l=1}^{N_{2}}e_{-1}(k_{j}-\lambda_{l})e_{5}(k_{j}-\lambda_{l})
\\ &&
\nonumber\prod_{n=1}^{\infty}\prod_{\alpha=1}^{M_{n}}e_{-(n-1)}(k_{j}-\Lambda_{\alpha}^{n})
\\ && \prod_{n=1}^{\infty}\prod_{\alpha=1}^{M_{n}}
e_{-(n+1)}(k_{j}-\Lambda_{\alpha}^{n}),
\end{eqnarray}
\begin{eqnarray}
\nonumber\exp(\mathrm{i}2\lambda_{j}L) &=&
\prod_{l=1}^{N_{1}}e_{-1}(\lambda_{j}-k_{l})e_{5}(\lambda_{j}-k_{l})
\\ && \prod_{l\neq j}^{N_{2}}
e_{-2}(\lambda_{j}-\lambda_{l})e_{4}(\lambda_{j}-\lambda_{l})e_{6}(\lambda_{j}-\lambda_{l}),
\end{eqnarray}
\begin{eqnarray}
\nonumber\lefteqn{\prod_{l=1}^{N_{1}}e_{(n-1)}(\Lambda_{\alpha}^{n}-k_{l})e_{(n+1)}(\Lambda_{\alpha}^{n}-k_{l})}
\\ &&
=-\prod_{m=1}^{\infty}\prod_{\beta=1}^{M_{m}}E_{mn}(\Lambda_{\alpha}^{n}-\Lambda_{\beta}^{m}),
\end{eqnarray}
where $E_{mn}(x)$ is defined as
\begin{equation}
E_{mn}(x)=\left\{\begin{array}{lll}
    e_{|m-n|}(x)e_{|m-n|+2}^{2}(x)\ldots \\
e_{m+n-2}^{2}(x)e_{m+n}(x), & \hbox{for $n\neq m$,} \\
    e_{2}^{2}(x)e_{4}^{2}(x)\ldots \\
e_{2n-2}^{2}(x)e_{2n}(x), & \hbox{for $n=m$.} \\
\end{array}\right.
\end{equation}


Following the technique pioneered by Yang and Yang \cite{Yang1969}
for spinless bosons, the logarithm of equations (A1) to (A3) then
gives
\begin{eqnarray}
\nonumber k_{j}L &=& 2\pi I_{j}-\sum_{l\neq
j}^{N_{1}}\theta\left(\frac{k_{j}-k_{l}}{4c'}\right)
\\ && \nonumber +\sum_{l=1}^{N_{2}}\left[\theta\left(\frac{k_{j}-\lambda_{l}}{c'}\right)
-\theta\left(\frac{k_{j}-\lambda_{l}}{5c'}\right)\right] \\ &&
+\sum_{n=1}^{\infty}\sum_{\alpha=1}^{M_{n}}\left[\theta\left(\frac{k_{j}-\Lambda_{\alpha}^{n}}{(n-1)c'}\right)
+\theta\left(\frac{k_{j}-\Lambda_{\alpha}^{n}}{(n+1)c'}\right)\right],
\end{eqnarray}
\begin{eqnarray}
\nonumber 2\lambda_{j}L &=& 2\pi
J_{j}+\sum_{l=1}^{N_{1}}\left[\theta\left(\frac{\lambda_{j}-k_{l}}{c'}\right)-\theta\left(\frac{\lambda_{j}-k_{l}}{5c'}\right)\right]
\\ && \nonumber +\sum_{l\neq
j}^{N_{2}}\left[\theta\left(\frac{\lambda_{j}-\lambda_{l}}{2c'}\right)-\theta\left(\frac{\lambda_{j}-\lambda_{l}}{4c'}\right)\right.
\\ &&
\left.-\theta\left(\frac{\lambda_{j}-\lambda_{l}}{6c'}\right)\right],
\end{eqnarray}
\begin{eqnarray}
\nonumber\lefteqn{\sum_{l=1}^{N_{1}}\left[\theta\left(\frac{\Lambda_{\alpha}^{n}-k_{l}}{(n-1)c'}\right)
+\theta\left(\frac{\Lambda_{\alpha}^{n}-k_{l}}{(n+1)c'}\right)\right]} \\
&& =2\pi
L_{\alpha}^{n}+\sum_{m=1}^{\infty}\sum_{\beta=1}^{M_{m}}\Theta_{mn}
\left(\frac{\lambda_{\alpha}^{n}-\Lambda_{\beta}^{m}}{c'}\right),
\end{eqnarray}
where $\theta(x)=2\tan^{-1}(x)$ and
\begin{equation}
\Theta_{mn}(x)=\left\{\begin{array}{lll}
    \theta\left(\frac{x}{|m-n|}\right)+2\theta\left(\frac{x}{|m-n|+2}\right)+\ldots
    \\
    +2\theta\left(\frac{x}{m+n-2}\right)+\theta\left(\frac{x}{m+n}\right), &  \hbox{for $n\neq m$,}
    \\
    2\theta\left(\frac{x}{2}\right)+2\theta\left(\frac{x}{4}\right)+\ldots \\ +
    2\theta\left(\frac{x}{2n-2}\right)+\theta\left(\frac{x}{2n}\right), & \hbox{for $n=m$.} \\
\end{array}\right.
\end{equation}

Writing the occupied distribution functions of the unpaired $k$'s,
the paired $k$'s and the $\Lambda$-strings as $\rho_{1}(k)$,
$\rho_{2}(k)$ and $\sigma_{n}(k)$ and their corresponding unoccupied
distribution functions as $\rho_{1}^{h}(k)$, $\rho_{2}^{h}(k)$ and
$\sigma_{n}^{h}(k)$, we take the thermodynamic limit of the above
equations to obtain the integral equations
\begin{eqnarray}
\nonumber \rho_{1}(k)+\rho_{1}^{h}(k) &=&
\frac{1}{2\pi}+a_{4}\ast\rho_{1}(k) \\
&& \nonumber +[a_{5}-a_{1}]\ast\rho_{2}(k)
\\ && -\sum_{n=1}^{\infty}[a_{n-1}+a_{n+1}]\ast\sigma_{n}(k),
\end{eqnarray}
\begin{eqnarray}
\nonumber \rho_{2}(k)+\rho^{h}_{2}(k) &=&
\frac{1}{\pi}+[a_{5}-a_{1}]\ast\rho_{1}(k)
\\ && +[a_{6}+a_{4}-a_{2}]\ast\rho_{2}(k),
\end{eqnarray}
\begin{eqnarray}
\nonumber \sigma_{n}(k)+\sigma_{n}(k) &=&
[a_{n-1}+a_{n+1}]\ast\rho_{1}(k)
\\ && -\sum_{m=1}^{\infty}T_{mn}\ast\sigma_{m}(k),
\end{eqnarray}
where
\begin{equation}
f\ast g(x)=\int_{-\infty}^{\infty}f(x-x')g(x')dx',
\label{Convolution}
\end{equation}
\begin{equation}
a_{n}(x)=\frac{1}{\pi}\frac{n|c'|}{(nc')^{2}+x^{2}}, \label{a}
\end{equation}
and
\begin{equation}
T_{mn}(x)=\left\{%
\begin{array}{ll}
    a_{|m-n|}(x)+2a_{|m-n|+2}(x)+\ldots \\ +2a_{m+n-2}(x)+a_{m+n}(x), & \hbox{for $n\neq m$,}
    \\
    2a_{2}(x)+2a_{4}(x)+\ldots \\ +2a_{2n-2}(x)+a_{2n}(x), & \hbox{for $n=m$.} \\
\end{array}%
\right.
\end{equation}

The distribution functions are related to the particle numbers via
the relations
\begin{eqnarray}
n_{1} &=& \frac{N_{1}}{L}=\int_{-\infty}^{\infty}\rho_{1}(k)dk, \\
n_{2} &=& \frac{N_{2}}{L}=\int_{-\infty}^{\infty}\rho_{2}(k)dk ,\\ m
&=& \sum_{n=1}^{\infty}n\int_{-\infty}^{\infty}\sigma_{n}(k)dk.
\end{eqnarray}
The total number of microstates in an interval $dk$ is
\begin{eqnarray}
\nonumber dW &=&
\frac{(L(\rho_{1}(k)+\rho_{1}^{h}(k))dk)!}{(L\rho_{1}(k)
dk)!(L\rho_{1}^{h}(k)dk)!}
\\ && \nonumber \times\frac{(L(\rho_{2}(k)+\rho_{2}^{h}(k))dk)!}{(L\rho_{2}(k)
dk)!(L\rho_{2}^{h}(k)dk)!}
\\ && \times\prod_{n=1}^{\infty}\frac{(L(\sigma_{n}(k)+\sigma_{n}^{h}(k))dk)!}{(L\sigma_{n}(k)
dk)!(L\sigma_{n}^{h}(k)dk)!}.
\end{eqnarray}
Through the use of Stirling's approximation, the entropy is written
as
\begin{eqnarray}
\nonumber\frac{S}{L} &=&
\int_{-\infty}^{\infty}\left[(\rho_{1}+\rho_{1}^{h})\ln(\rho_{1}+\rho_{1}^{h})\right.
\\ && \nonumber \left.-\rho_{1}\ln\rho_{1}-\rho_{1}^{h}\ln\rho_{1}^{h}\right]dk
\\ && \nonumber +\int_{-\infty}^{\infty}\left[(\rho_{2}+\rho_{2}^{h})\ln(\rho_{2}
+\rho_{2}^{h})\right.
\\ && \nonumber \left.-\rho_{2}\ln\rho_{2}-\rho_{2}^{h}\ln\rho_{2}^{h}\right]dk
\\ && \nonumber
+\sum_{n=1}^{\infty}\int_{-\infty}^{\infty}\left[(\sigma_{n}+\sigma_{n}^{h})\ln(\sigma_{n}+\sigma_{n}^{h})\right.
\\ &&
\left.-\sigma_{n}\ln\sigma_{n}-\sigma_{n}^{h}\ln\sigma_{n}^{h}\right]dk.
\end{eqnarray}

The Gibbs free energy per unit length is
\begin{equation}
\Omega=\frac{E}{L}-\mu n-\frac{TS}{L}+\frac{E_{z}}{L}
\end{equation}
where $\mu$ is the chemical potential, the energy per unit length is
\begin{equation}
\frac{E}{L}=\int_{-\infty}^{\infty}
k^{2}\rho_{1}(k)dk+\int_{-\infty}^{\infty}(2k^{2}-2c'^{2})\rho_{2}(k)dk
\end{equation}
and the Zeeman energy per unit length is given in terms of the
external magnetic field $H$
\begin{equation}
\frac{E_{z}}{L}=-H\int_{-\infty}^{\infty}\rho_{1}(k)dk+H\sum_{n=1}^{\infty}n\int_{-\infty}^{\infty}\sigma_{n}(k)dk.
\end{equation}
Minimizing the Gibbs free energy and going through a similar
procedure as shown in \cite{Takahashi1999}, we arrive finally at the
TBA equations (\ref{TBA}).

\clearpage

\begin{figure}[ht]
\includegraphics[width=.8\columnwidth]{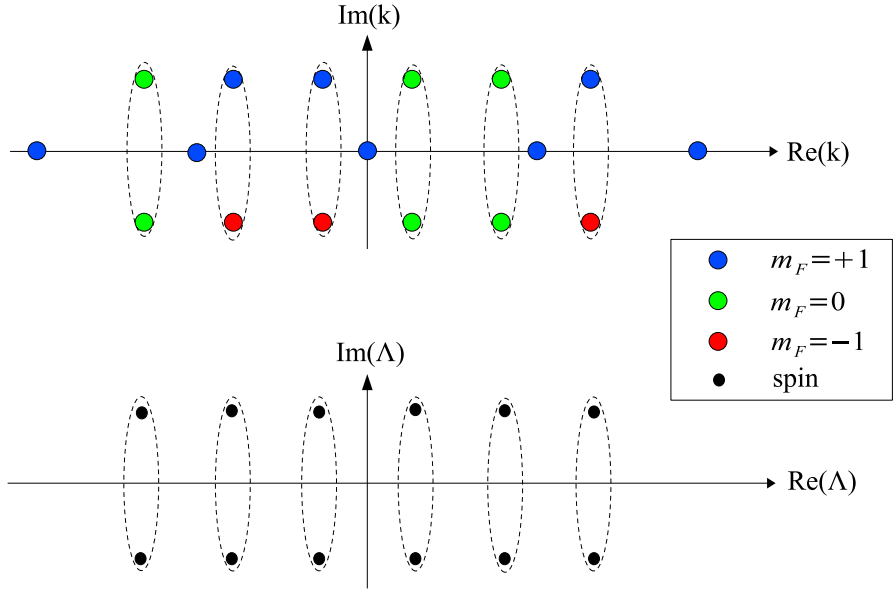}
\caption{(color online) Schematic configuration of the quasimomenta
$k$ and spin rapidities $\Lambda$ in the complex plane for the
ground state in the strong coupling limit with $N=17$ and $M=12$.
Dashed boundaries indicate pair formations with the same real parts.
Each pair in $k$-space has a corresponding pair in $\Lambda$-space.
As mentioned in the text, the individual particle numbers for
$m_{F}=0,\pm1$ are not conserved and thus can fluctuate.}
\end{figure}

\begin{figure}[ht]
\includegraphics[width=.8\columnwidth]{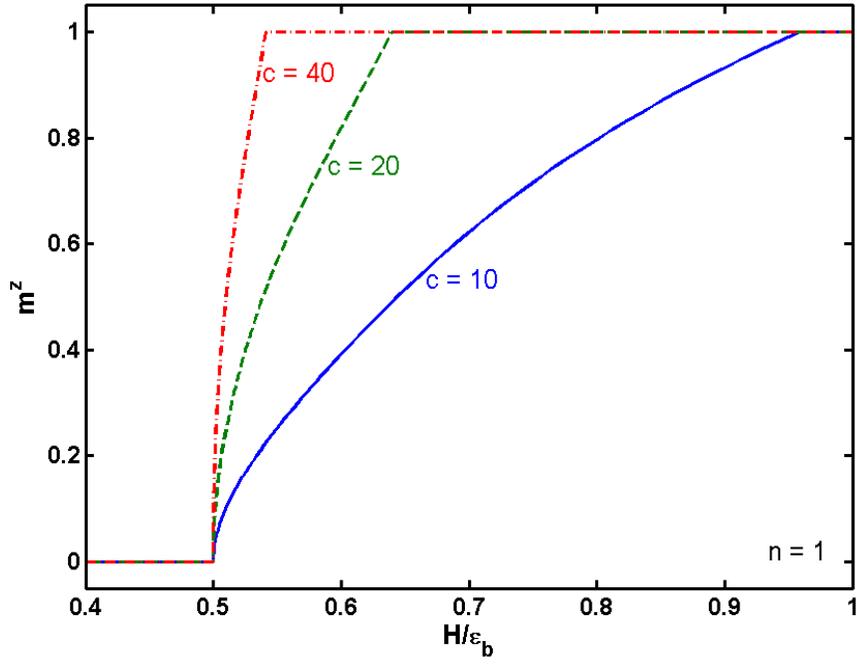}
\caption{Normalized magnetization $m^z$ versus re-scaled magnetic field $H/\epsilon_{b}$ for different values of interaction strength
$c$.}
\end{figure}

\begin{figure}[ht]
\includegraphics[width=.8\columnwidth]{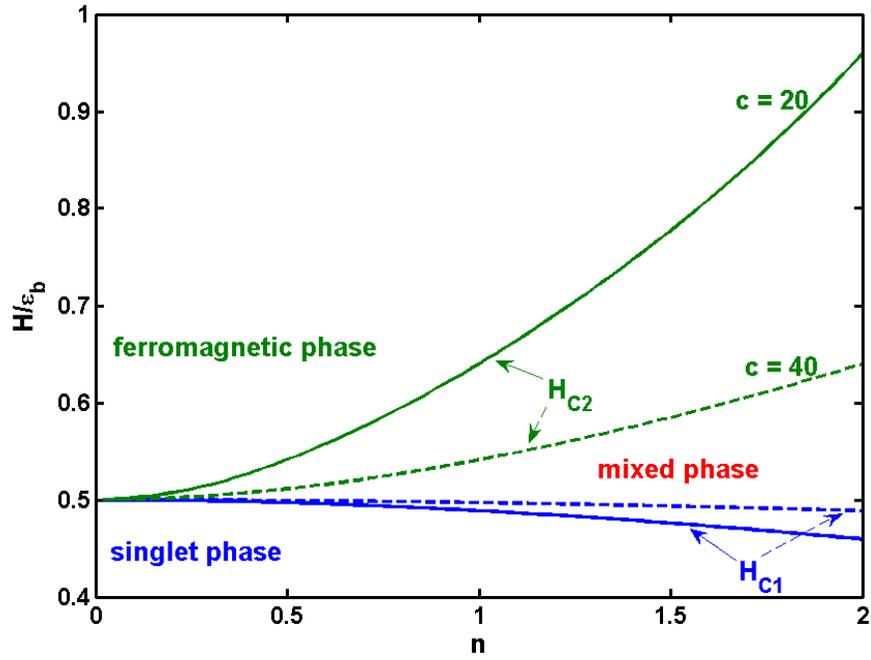}
\caption{Phase diagram in plane $(n, H)$ for different values of interaction strength
$c$.}
\end{figure}

\end{document}